\pdfoutput=1

\documentclass[aps,prl,amsmath,amssymb,
showpacs,
reprint,%
]{revtex4-1}

\usepackage{graphicx}
\usepackage{dcolumn}
\usepackage{bm}

\usepackage{color}

\begin{document}

\preprint{}

\title{Two-dimensional transition metal dichalcogenides under electron
  irradiation: defect production and doping}

\author{Hannu-Pekka Komsa$^1$}
\author{Jani Kotakoski$^{1,2}$}
\author{Simon Kurasch$^3$}
\author{Ossi Lehtinen$^1$}
\author{Ute Kaiser$^3$}
\author{Arkady V. Krasheninnikov$^{1,4}$}
\affiliation{
$^1$Department of Physics, University of Helsinki,
P.O. Box 43, 00014 Helsinki, Finland
}
\affiliation{
$^2$
Department of Physics, University of Vienna, 
Boltzmanngasse 5, 1190 Wien, Austria
}
\affiliation{
$^3$ Central Facility for Electron Microscopy, Group of Electron Microscopy of
Materials Science, University of Ulm, 89081 Ulm, Germany }
\affiliation{
$^4$Department of Applied Physics, Aalto University,
P.O. Box 11100, 00076 Aalto, Finland
}

\date{\today}

\begin{abstract}
Using first-principles atomistic simulations, we study the response of
atomically-thin layers of transition metal dichalcogenides (TMDs) -- a
new class of two-dimensional inorganic materials with unique
electronic properties -- to electron irradiation.
We calculate displacement threshold energies for atoms in 21 different
compounds and estimate the corresponding electron energies required to
produce defects. For a representative structure of MoS$_2$, we carry out
high-resolution transmission electron microscopy experiments 
and validate our theoretical predictions via observations of
vacancy formation under exposure to a 80~keV electron beam.  We
further show that TMDs can be doped by filling the vacancies created
by the electron beam with impurity atoms. Thereby, our results not
only shed light on the radiation response of a system with reduced
dimensionality, but also suggest new ways for engineering the
electronic structure of TMDs.
\end{abstract}



\maketitle


Isolation of a single sheet of graphene in 2004 \cite{Novoselov04sci}
indicated that strictly two-dimensional (2D) materials can exist
at finite temperatures.
Indeed, inorganic 2D systems such as individual hexagonal BN and transition
metal dichalcogenide (TMD) layers were later on manufactured by 
mechanical \cite{Novoselov2005pnas,Radisavljevic2011} and
chemical \cite{Eda2011,Coleman11_Sci} exfoliation of their layered bulk
counterparts, as well as by chemical vapor deposition
\cite{Kim12cvd,Zhan2012sm}. 
Recently, TMDs with a common structural formula MeX$_2$, where Me stands for 
transition metals (Mo, W, Ti, etc.) and X for chalcogens (S, Se, Te), 
have received considerable attention.
These 2D materials are expected to have electronic properties varying 
from metals to wide-gap semiconductors, similar to their bulk 
counterparts \cite{Wilson69_AP,Ataca12},
and excellent mechanical characteristics \cite{Castellanos-Gomez2012}.
The monolayer TMD materials have already shown a good potential in 
nanoelectronic \cite{Radisavljevic2011,Li2012sm,Popov12_PRL} and
photonic \cite{Mak10_PRL,Yin2012acs,Eda2011} applications.

Characterization of the
$h$-BN~\cite{MeyerNL09-BN,JinPRL09-BN,Meyer11NM} and
TMD~\cite{Kim12cvd,Coleman11_Sci,Brivio2011nl} samples has 
extensively been carried out using high-resolution transmission electron
microscopy (HR-TEM). During imaging, however, energetic electrons in
the TEM can give rise to production of defects due to ballistic
displacements of atoms from the sample and beam-stimulated chemical
etching~\cite{Molhave2007},
as studies on {\it h}-BN membranes also indicate
~\cite{MeyerNL09-BN,JinPRL09-BN,Kotakoski10BN,Meyer11NM}.

Contrary to $h$-BN, very
little is known about the effects of electron irradiation on TMDs. So
far, atomic defects have been observed via HR-TEM in WS$_2$
nanoribbons encapsulated inside carbon nanotubes at electron
acceleration voltage of 60~kV~\cite{Liu2011nc} as well as at the edges
of MoS$_2$ clusters under 80~kV irradiation~\cite{Hansen11_ACIE},
while no significant damage or amorphization was reported for MoS$_2$
sheets at 200~kV~\cite{Brivio2011nl} -- a surprising result taking
into account the relatively low atomic mass of the S atom.
Clearly, precise microscopic knowledge of defect production 
in TMDs under electron irradiation is highly desirable for assessing 
the effects of the beam on the samples. This
knowledge would allow designing experimental conditions required
to minimize damage, as well as developing beam-mediated post-synthesis doping
techniques.  Moreover, information on the displacement thresholds is
important in the context of fundamental aspects of the interaction of
beams of energetic particles with solids, as the reduced
dimensionality may give rise to an irradiation response different from
that in the bulk counterpart of the 2D
material~\cite{Kra10JAP}.

Here, by employing first-principles simulations, we study the behavior
of a representative number of TMDs (21 compounds) under electron
irradiation, and calculate the threshold energies for atomic
displacements in each system, as well as displacement cross sections
as functions of electron beam energy. In the case of MoS$_2$, we also
carry out HR-TEM experiments and provide evidence of
electron-irradiation-induced production of vacancies in this
material. In addition -- inspired by the recent advances in introducing
impurities in  $h$-BN monolayers \cite{Wei11dop,Krivanek10nat}
-- we discuss irradiation-mediated doping of TMD materials.

For all calculations in this work, we rely on the density-functional
theory (DFT) with the PBE exchange-correlation functional
~\cite{Perdew96} and the projector augmented wave
formalism as implemented in the simulation package
VASP~\cite{kres1,kres2}. In order to obtain a comprehensive picture of
the irradiation response of TMDs, we consider a large set of layered
TMDs: MoX$_2$, WX$_2$, NbX$_2$, TaX$_2$,
PtX$_2$, TiX$_2$, and VX$_2$ (where X=S, Se, or Te), which have
similar crystal structures. 

\begin{figure}[!ht]
\begin{center}
  \includegraphics[width=8.5cm]{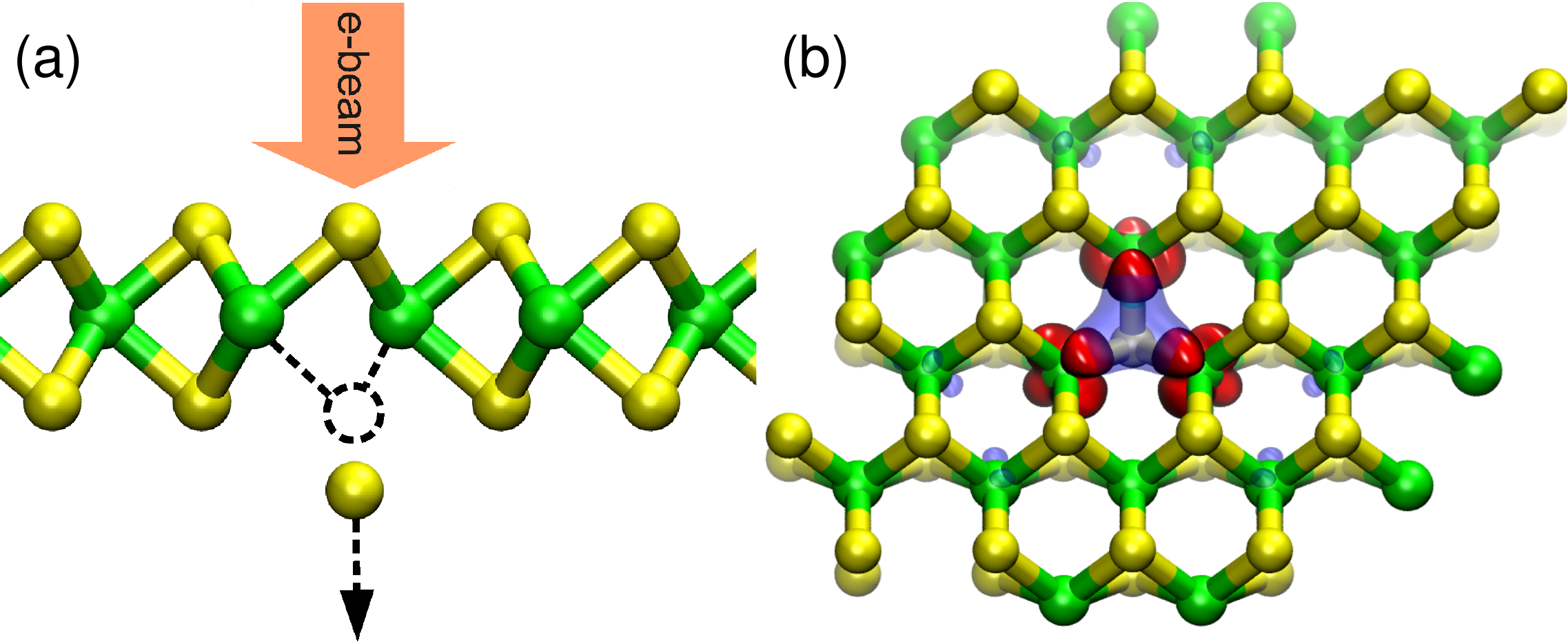}
\end{center}
\caption{\label{fig:setup}
(Color online)  
(a) The setup used in the dynamical DFT simulations of atom sputtering
from TMDs under electron irradiation. Initial energy
acquired due to the impact of an energetic electron was assigned to 
the recoil atom, then DFT molecular dynamics was used to model the 
evolution of the system.  (b) MoS$_2$ sheet with an S vacancy.  The charge
densities of the occupied and unoccupied defect states are visualized by
(blue) transparent and (red) solid isosurfaces, respectively.
}
\end{figure}

We started our study by calculating the displacement threshold energy $T_d$ 
(the minimum initial kinetic energy of the recoil atom) for sputtering an
atom from the material. As in our previous simulations 
for graphene~\cite{Kotakoski12acsn} and BN~\cite{Kotakoski10BN}
monolayers, an initial velocity was assigned to the recoil atom
(corresponding to instantaneous momentum transfer from the electron to
the atom during the impact), then DFT molecular dynamics was used to
model the time evolution of the system. In practice, the initial
kinetic energy of the recoil atom was increased until it was high
enough for the atom to be displaced from its lattice site without an
immediate recombination with the resulting vacancy.
The calculations were carried out using a $5\times5$
supercell of a MeX$_2$ monolayer. Test simulations for larger systems
gave essentially the same results. The atomic structure of a MoS$_2$
layer and the simulation setup are shown in Fig.~\ref{fig:setup}(a).

\begin{figure}[!ht]
\begin{center}
  \includegraphics[width=8.5cm]{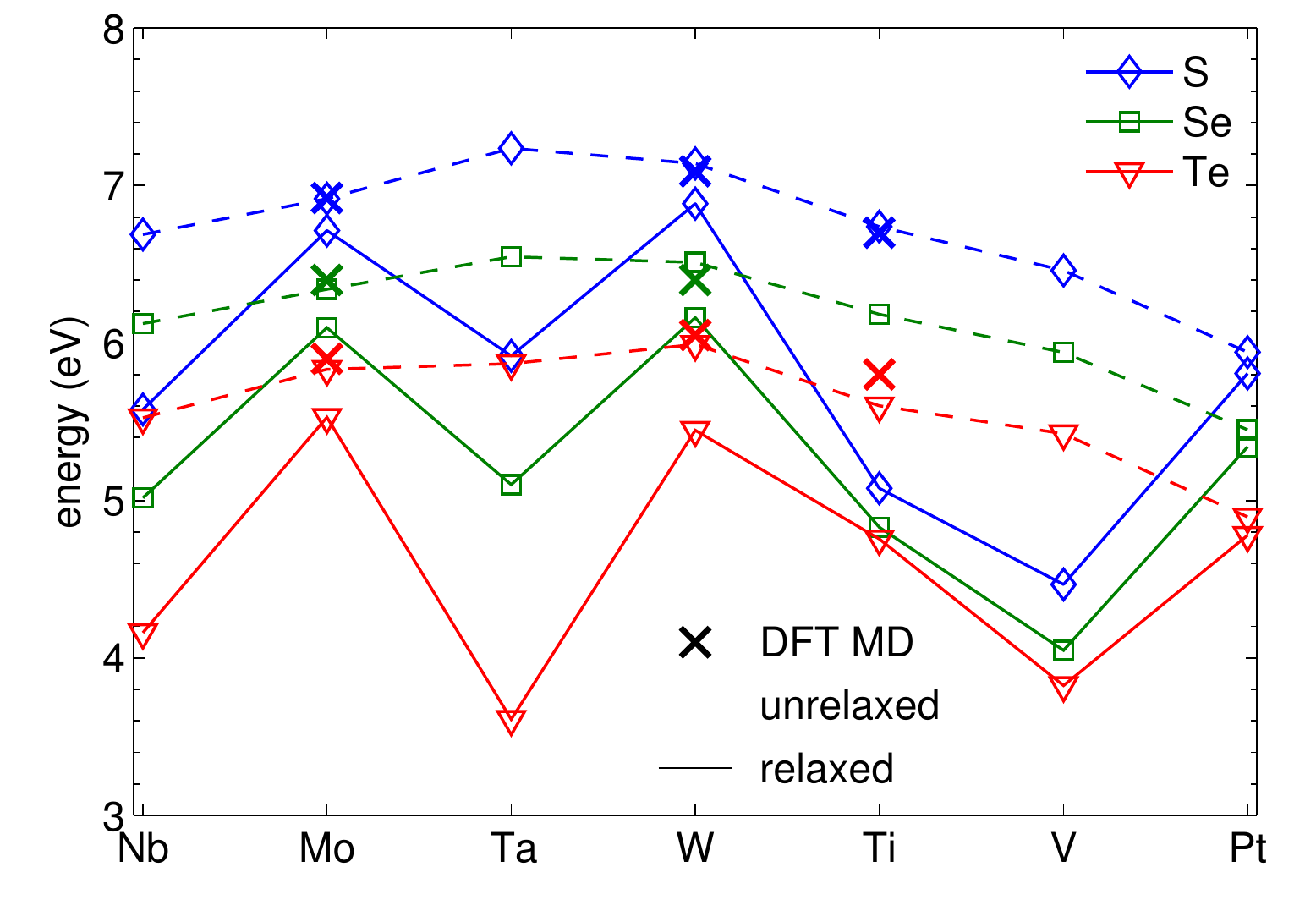}
\end{center}
\caption{\label{fig:Eform}
(Color online)
Displacement threshold energies $T_d$ obtained from DFT molecular dynamics 
calculations (crosses) and formation energies of chalcogen vacancies 
with non-relaxed (dashed lines) and relaxed (solid lines) geometries 
in transition metal dichalcogenides MeX$_2$, X=S, Se, Te.
}
\end{figure}

$T_d$ required for displacing a chalcogen atom from the bottom
layer of the sheet [cf. Fig.~\ref{fig:setup}(a)]
are presented in
Fig.~\ref{fig:Eform}. In addition to the prototypical MoX$_2$ and
WX$_2$, we also present results for TiS$_2$ and TiTe$_2$. 
As evident from the figure, $T_d \in [5,7]$ eV for all studied compounds.

We also calculated the vacancy formation energies ($E_f$) for each of
the compounds to see how it correlates with $T_d$. We defined $E_f$ as:
\begin{equation}
E_f = E_{\rm vac}-(E_{\rm bulk}-\mu_\mathrm{X}),
\end{equation}
where $E_{\rm bulk}$ and $E_{\rm vac}$ are the energies of the
pristine and vacancy containing supercells, respectively.  The
chemical potential $\mu_\mathrm{X}$ of the chalcogen species is taken
as the energy of the isolated atom to enable a straightforward
comparison with the results of dynamical simulations. $E_f$, with and
without relaxation of the atomic structure of the layer, is also presented
in Fig.~\ref{fig:Eform}.

In the non-relaxed case, the energetics is very similar for all
materials, and -- as can  readily be noticed -- the agreement between
$T_d$ and the {\it non-relaxed} $E_f$ is striking. This is
because during the sputtering of chalcogen atoms from the outermost
layer, little energy is transmitted to the surrounding metal atoms due
to sufficiently fast sputtering event and the rigidity of the structure. 
The energies for the relaxed geometries show a more
intriguing behavior. Atomic relaxation for some systems evidently
gives rise to a considerable drop in $E_f$, so that the similarity
to $T_d$ is lost. This drop quantifies the degree of structural
relaxation around the vacancy, which
is minor for MoS$_2$, see Fig.~\ref{fig:setup}(b).
The analysis of the electronic structure 
revealed an occupied bonding type vacancy state close to
valence band maximum and an empty anti-bonding type state in the mid
gap,
which stabilizes the structure. For the occupied bonding defect state,
the electronic charge is localized at the vacancy site,
analogous to bulk MoS$_2$, where Mo atoms donate electrons to S atoms. 
This is true for all of the semiconducting materials: MoX$_2$,
WX$_2$, and PtX$_2$.  The rest of the considered materials are metals
or semimetals, for which the bonding vacancy state may become
unoccupied, which is reflected in larger relaxation and lower
formation energies.

\begin{figure}[!ht]
\begin{center}
  \includegraphics[width=8.5cm]{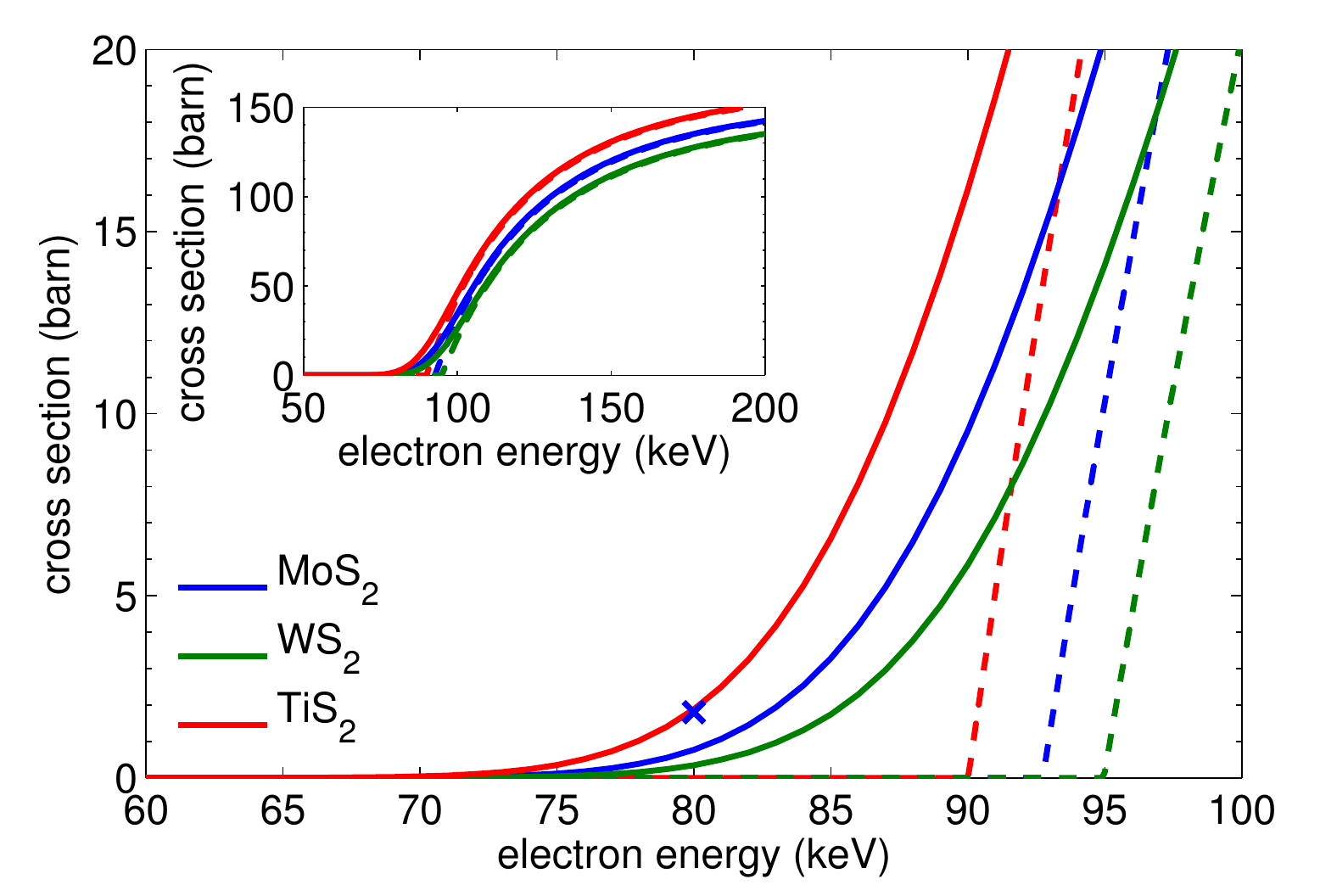}
\end{center}
\caption{\label{fig:crosect} (Color online) Cross section for
  sputtering a sulfur atom from MoS$_2$, WS$_2$, and TiS$_2$ sheets as
  calculated through the McKinley-Feshbach formalism and the dynamical
  values of the displacement thresholds. Dotted lines are the data for
  the static lattice, and solids lines are the results of calculations
  where lattice vibrations are taken into account assuming a
  Maxwell-Boltzmann velocity distribution. The cross denotes the
  experimentally determined cross section for MoS$_2$. The inset shows the same
  data for a larger range of electron energies.}
\end{figure}

Knowing $T_d$, it is possible to estimate the electron threshold
energy through the relativistic binary collision formula and the atom
displacement cross section (for relatively light atoms) by using the
McKinley-Feshbach formalism~\cite{McKinley48}. 
$E_f$ for Se and Te compounds are smaller than in S compounds, but
-- due to the higher atomic mass -- their creation through ballistic electron
impacts requires significantly higher electron energies.  In the case
of MoS$_2$, MoSe$_2$, and MoTe$_2$, $T_d$ of 6.9, 6.4 and 5.9 eV
correspond to electron energies of about 90, 190, and 270 keV, as
calculated assuming a static lattice.  For other compounds the
required electron energies should be of similar magnitude, based on
the close values of $T_d$ in Fig.~\ref{fig:Eform}. 
An accurate estimation of the displacement cross section
requires including the effects of lattice vibrations on the energy
transferred from an electron to a target atom~\cite{Meyer2012PRL}. 
We calculated the cross sections for vacancy production as a function 
of electron energy for MoS$_2$, WS$_2$, and TiS$_2$ beyond the static 
lattice approximation, as shown in Fig.~\ref{fig:crosect}. 
We stress that the production of S vacancies for practically all 
TMDs is within the energies commonly used in TEM studies.

The displacement thresholds for chalcogen atoms in the (top) layer
facing the beam proved to be considerably higher than for the bottom
chalcogen layer, as the displaced atom is ``stopped'' by the other
layers. However, after a vacancy is created in the bottom layer, the
threshold energy for the top S atom in MoS$_2$ to be displaced and
fill the vacancy is about 8.1~eV. This is similar in magnitude to the
threshold for displacing S atom from the bottom layer (6.9~eV), and
thus formation of vacancy columns should be possible even at 80~kV
when lattice vibrations are accounted for. $T_d$
for transition metals are even higher, since they are bonded to six
neighbors and similarly stopped by the S layer. For instance, about 
20~eV is required to displace Mo atom from its site in the MoS$_2$
lattice, which corresponds to electron energy of 560~keV.
Naturally, under such conditions the S sublattice is quickly destroyed. 
Formation of transition metal vacancies is thus considered highly unlikely.

With regard to possible vacancy agglomeration under continuous
irradiation, we found that creation of a vacancy does not alter the
formation energy in the neighboring sites in the semiconducting TMDs.
Thus, we do not expect accelerated formation of large vacancy clusters.
In the same vein, however, it is worth noting that chalcogen atoms may
also be sputtered fairly easily from the edges of
nanostructures~\cite{Liu2011nc,Hansen11_ACIE}. For example, our
calculations for a WS$_2$ ribbon show that the chalcogen atoms at the
edge can have a displacement threshold as low as 4.2 eV, as compared
to 7.0 eV away from the edge.

To check our theoretical results on irradiation-induced vacancy
formation in MoS$_2$, we experimentally studied the evolution of a
MoS$_2$ sheet under an 80~keV electron beam. First, free standing
single layer MoS$_2$ samples were prepared by mechanical exfoliation
of natural MoS$_2$ bulk crystals, followed by characterization via
optical microscopy on a Si+90nm SiO$_2$ substrate and transfer to a
perforated TEM support film (Quantifoil),
similar to graphene samples
~\cite{Meyer2008b}. 
The TEM grid was adhered to the SiO$_2$
surface by evaporating isopropanol on top of it. After this, the
silica was etched with KOH. Aberration-corrected (AC) HRTEM imaging 
was carried out in an image-side Cs
corrected FEI TITAN microscope at a primary beam energy of 80~keV.
The contrast difference between the Mo and S sublattice is clearly
detectable in the AC-HRTEM images proving the
single layer nature of the sheet (for a double layer the contrast
would be identical as Mo is stacked above S).  This is also confirmed
in diffraction measurements, as successive diffraction spots from one
$\{hkl\}$ family show different intensity, whereas for bi- and
multilayers they are equal~\cite{Brivio2011nl}. The analyzed intensity
ratio of the $\{\bar{1}100\}$ diffraction spots was found to be
$1.07$.

\begin{figure}[ht!]
\begin{center}
\includegraphics[width=0.95\linewidth]{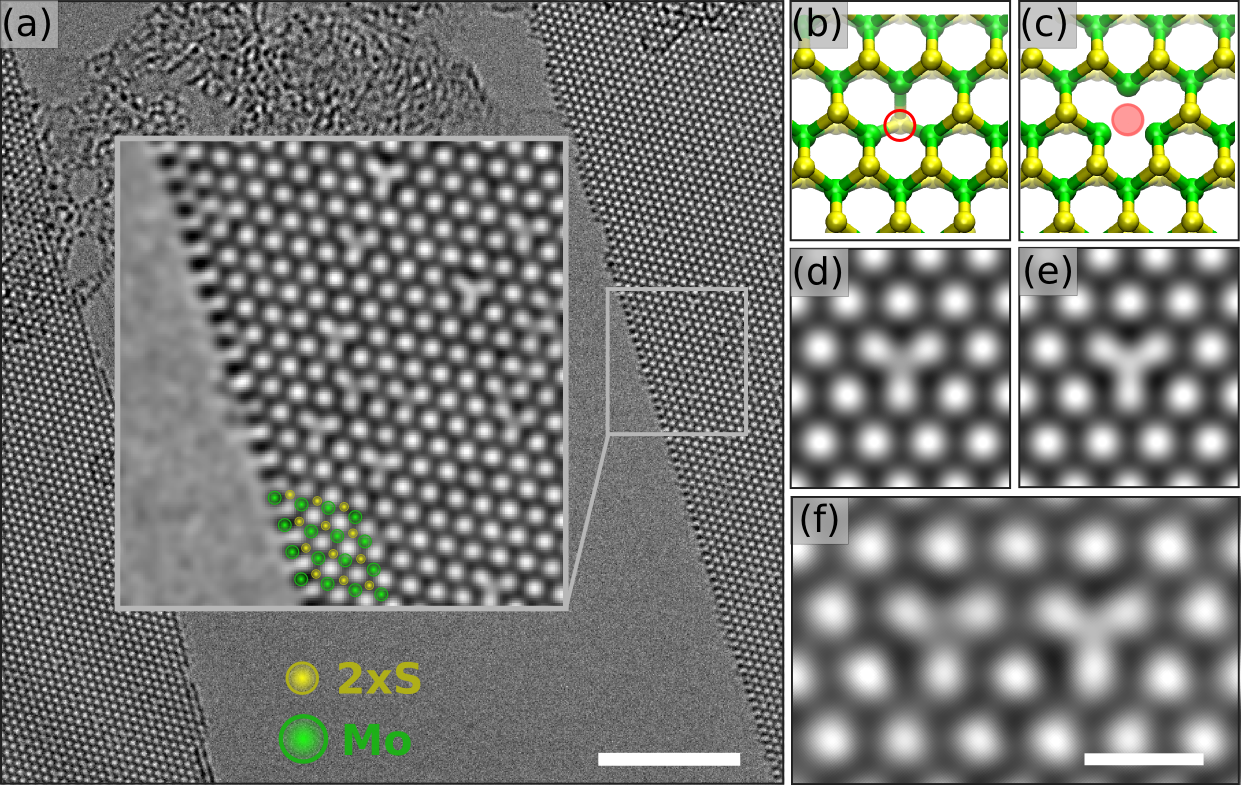}
\end{center}
\caption{\label{fig:exp} (Color online) AC-HRTEM images of
  single-layer MoS$_2$. Atoms are dark and white spots correspond
  to the holes in the hexagonal structure.
  During continuous 80~keV electron irradiation
  a crack formed in the membrane (a). The edges are terminated with Mo
  atoms, as can be determined from the contrast in the inset.  Also an
  increasing number of S vancancies is observed. Most of them are
  single vacancies but also double vacancies, where the top and bottom
  S atoms are removed, can be found. Structure models and
  corresponding HRTEM image simulations for both vacancy types are
  shown in (b, d) for the single- and (c, e) for the double
  vacancy. Experimental examples are shown in (f). To enhance the
  visibility of the defect, a Gaussian filter (0.7 \AA\ FWHM) was
  applied to (d, e) and (f). Scale bars are 5 nm (a) and 5 \AA\ (f).}
\end{figure}

During continuous imaging we found an increasing number of vacancy
sites (exclusively on the S sublattice) accompanied by crack formation
[see Fig.~\ref{fig:exp}(a)] and lateral shrinkage of the membrane.
Counting the actual number of sputtered atoms as
in Ref.~\onlinecite{Meyer2012PRL}, the cross section for sputtering
was found to be $1.8$ barn, which is in a reasonable agreement with
the calculated cross-section of $0.8$ barn, taking into account that
the theoretical estimates are very sensitive to inaccuracies in the
parameters of the model (e.g., $T_d$ and the velocity distribution) at
energies below the static threshold.

In Fig.~\ref{fig:exp}(d,e) we present simulated TEM images \cite{KochThesis}
for the single and double vacancies, respectively, based on atomic structures 
(Fig. \ref{fig:exp}(b,c)) obtained from the DFT calculations.
Similar defects are observed in the experimental TEM images 
[Fig.~\ref{fig:exp}(f)].
Different defects can clearly be distinguished by analyzing the (Michelson)
contrast relative to the contrast of the Mo atoms in the pristine area.
We find that the experimental (simulated) ratios are 0.9 (0.9) for a sulfur 
column, 0.5 (0.4) for the single- and 0.2 (0.2) for the double vacancy. 

\begin{figure}[!ht]
\begin{center}
  \includegraphics[width=8.5cm]{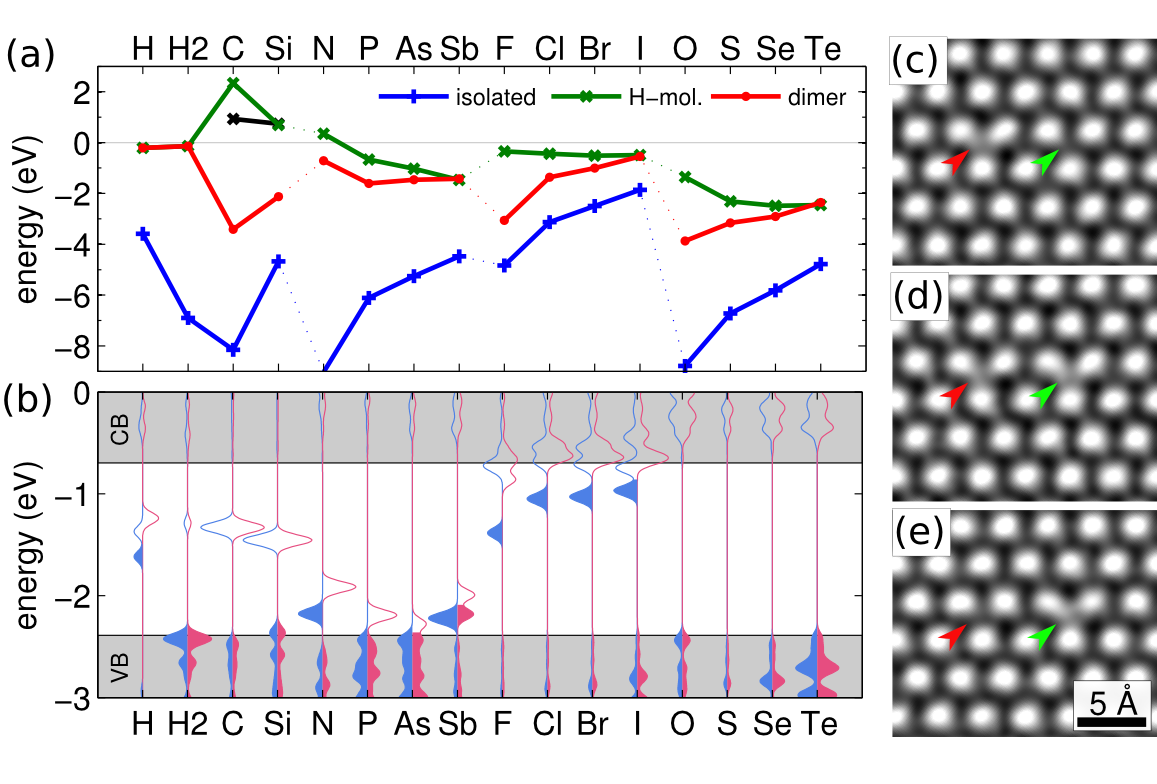}
\end{center}
\caption{\label{fig:sub} (Color online) (a) Substitution energies of
  the impurity atoms in MoS$_2$ layer calculated for three different
  chemical potentials: isolated atoms (blue), diatomic molecules
  (red), and molecules with hydrogen (green).  The solid phase
  reference is also given for carbon and silicon (black).  (b)
  Respective local density of states of the impurity atoms, the
  spin-up and spin-down components. The gray areas denote the band
  edges of the MoS$_2$ layer. (c--e) Series of AC-HRTEM images
  demonstrating vacancy filling. The red arrow highlights an initial
  S vacancy that picks up an atom between (d) and (e),
  and the green arrow indicates an S atom that is
  sputtered away between (c) and (d), forming a single vacancy.
}

\end{figure}

Having shown that vacancies can be created in TMDs under electron
irradiation, we move on to study whether they could be consecutively
filled with other atomic species deliberately introduced into the TEM
chamber. We calculate the formation energy of substitutional defects
in MoS$_2$ and consider donors F, Cl, Br, and I; acceptors, N, P, As,
and Sb; double acceptors C and Si; hydrogen H and H$_2$; and
isoelectronic species O, S, Se, and Te.  The formation energies and
the local density of states (LDOS) around the substitution site are
shown in Fig.~\ref{fig:sub}.  We list the formation energies with
three different chemical potentials of the substituted (impurity)
species: the isolated atoms, diatomic molecules, or molecules with
hydrogen (CH$_4$, SiH$_4$, NH$_3$, PH$_3$, AsH$_3$, HF, HCl, and HBr)
where we set $\mu_\mathrm{H} = \frac{1}{2}E_{{\rm H}_2}$.

Due to the high formation energy of the vacancy, all substitutions are
energetically favored with respect to the isolated atom.  However,
with respect to $\mu$ in the molecule, C, Si, and N substitution have
positive formation energies.  Obviously, even if the equilibrium
energetics does not favor the formation of the substitutional defect,
the substitution may still be achieved under electron beam, because
molecules like N$_2$ or hydrocarbons will constantly break apart under
the electron beam.  Thus post-synthesis electron-mediated doping may
also be realized in 2D TMDs, similar to BN
sheets~\cite{Wei11dop,Krivanek10nat}.
The LDOS shows that N, P, As, and Sb behave as acceptors, 
whereas F, Cl, Br, and I are likely to be donors.
C and Si have levels in the middle of the gap and the
isoelectronic species like O, Se or Te do not produce any localized
states, as expected.


Filling of the vacancies was also observed in the TEM 
images, as shown in the series of panels in Fig.~\ref{fig:sub}(c--e).
Although we could not identify the type of the impurity,
this example proves that electron-beam-mediated doping is possible. 
Consequently, through control of atomic species in the TEM chamber and the 
choice of the electron energy, modification of the physical properties 
of TMDs via electron beam should be attainable.

To conclude, we calculated atom displacement threshold energies in a number
of TMDs. These energies are a measure of the radiation hardness of the
material, and serve as critical input parameters in the Kinchin-Pease
and other semiclassical theories of defect production and ion stopping
\cite{Kra10JAP,Kinchin55}.
Here we use them to calculate electron displacement energies and
corresponding sputtering cross sections to quantitatively assess the
amount of damage created in 2D TMD materials during a TEM experiment
via knock-on processes.  Observations of vacancies in our experimental
AC-HRTEM images of single MoS$_2$ sheets validate our theoretical
predictions.  Finally, we observe filling of the vacancies and discuss
the prospects for electron-beam mediated doping of TMDs.

We thank R.~M.~Nieminen for fruitful discussions.  We acknowledge
financial support by the University of Helsinki Funds and the
Academy of Finland through several projects. SK and UK are grateful
to the German Research Foundation (DFG) and the State Baden Wuerttemberg 
in the frame of the SALVE
project and to the DFG in the frame of the SFB-TRR21
project. We also thank CSC Finland for generous grants of computer time.


%

\end{document}